\documentclass[aps,prd,twocolumn, preprintnumbers, floatfix, superscriptaddress,showpacs]{revtex4}
\usepackage[dvips]{graphics}
\usepackage{amsmath,amsfonts,bm}
\usepackage{epsfig,bm}
\usepackage{graphicx}
\def\MeijG[#1][#2][#3][#4][#5][#6]{G^{#1}_{#2}\left(#3,#4\left|\begin{array}{c}#5\\#6\end{array}\right|\right)}

\begin{document}

\preprint{}

\affiliation{Dipartimento di Scienze Fisiche,
             Universit\'a di Napoli "Federico II", Via Cintia, 80126 Napoli, Italia}
\affiliation{INFN-Sezione di Napoli, Via Cintia, 80126 Napoli, Italia}
\affiliation{Departament de F\'isica Te\`orica and IFIC, Universitat de Val\`encia-CSIC,
Apt. Correus 22085, E-46071 Val\`encia, Spain}       
\affiliation{INFN, Laboratori Nazionali di Frascati, 00044 Frascati, Italia}

\author{Luigi Cappiello}
\affiliation{Dipartimento di Scienze Fisiche,
             Universit\'a di Napoli "Federico II", Via Cintia, 80126 Napoli, Italia}
\affiliation{INFN-Sezione di Napoli, Via Cintia, 80126 Napoli, Italia}

\author{Oscar Cat\`a}
\affiliation{Departament de F\'isica Te\`orica and IFIC, Universitat de Val\`encia-CSIC,
Apt. Correus 22085, E-46071 Val\`encia, Spain}       
\affiliation{INFN, Laboratori Nazionali di Frascati, 00044 Frascati, Italia}

\author{Giancarlo D'Ambrosio}
\affiliation{INFN-Sezione di Napoli, Via Cintia, 80126 Napoli, Italia}

\title{Antisymmetric tensors in holographic approaches to QCD}

\begin{abstract}
We study real (massive) antisymmetric tensors of rank two in holographic models of QCD based on the gauge/string duality. Our aim is to understand in detail how the AdS/CFT correspondence describes correlators with tensor currents in QCD. To this end we study a set of bootstrapped correlators with spin-1 vector and tensor currents, imposing matching to QCD at the partonic level. We show that a consistent description of this set of correlators yields a very predictive picture. For instance, it imposes strong constraints on infrared boundary conditions and precludes the introduction of dilatonic backgrounds as a mechanism to achieve linear confinement. Additionally, correlators with tensor currents turn out to be especially sensitive to chiral symmetry breaking, thus offering an ideal testing ground for genuine QCD effects. Several phenomenological consequences are explored, such as the nontrivial interplay between $1^{+-}$ states and conventional $1^{--}$ vector mesons.  
\end{abstract}

\maketitle


\section{Introduction}

The original AdS/CFT conjecture by Maldacena~\cite{Maldacena:1997re} has played a pivotal role to study strongly coupled systems through their duality with supergravity theories at a large number of colors. 
In recent years there have been attempts to extend the gauge/string duality to incorporate genuine QCD effects such as confinement and chiral symmetry breaking. Incorporating such effects is subject to a certain degree of arbitrariness, and the AdS/CFT correspondence can only be used as a guiding principle. However, the phenomenological success of such models is quite remarkable, especially with vector mesons~\cite{Erlich:2005qh}.

In this paper we want to extend this analysis to antisymmetric tensors of rank two. There are two main motivations to pursue this program. The first comes from QCD: while $J_{\mu}={\bar{q}}\gamma_{\mu}q$ is the paradigmatic current to create spin-1 vector mesons, the antisymmetric tensor current $J_{\mu\nu}={\bar{q}}\sigma_{\mu\nu}q$ is also known to generate spin-1 states, not only the conventional $1^{--}$ vector mesons but also the more exotic $1^{+-}$ mesons. Therefore, the study of antisymmetric currents not only provides a description of $1^{+-}$ mesons, but also involves a nontrivial interplay with $1^{--}$ mesons. Such interplay has been studied in QCD by a simultaneous analysis of the two-point correlators $\Pi_{VV}$, $\Pi_{VT}$ and $\Pi_{TT}$~\cite{Craigie:1981jx,Cata:2008zc}. This simultaneous analysis is highly constraining, and we would like to understand whether simple holographic settings with the AdS/CFT correspondence can offer a satisfactory description of these correlators.  

A more formal motivation comes from the gravity side. The spectrum of compactified $d=10$ type-II B supergravity on AdS$_5$ typically contains massive rank two antisymmetric tensors~\cite{Kim:1985ez}. Massive 2-forms have already been studied in the context of the AdS/CFT correspondence~\cite{Arutyunov:1998xt,l'Yi:1998eu}. In this work we want to examine their role in holographic models of QCD.

According to the AdS/CFT dictionary, $p$-forms with five-dimensional masses $m$ coupled to currents with conformal dimension $\Delta$ satisfy
\begin{equation}\label{ads}
m^2=(\Delta-p)(\Delta+p-d)~.
\end{equation}
Following the AdS/CFT prescription for correlators developed in~\cite{Gubser:1998bc, Witten:1998qj}, the on-shell fields on the gravitational side are to be identified with the sources of the gauge theory currents. However, in order to bring the AdS/CFT correspondence closer to QCD one needs further ingredients, {\it{e.g.}}, fields in the fundamental representation and a mechanism for chiral symmetry breaking. The former can be accomplished through embeddings of probe branes, as shown in~\cite{Karch:2002sh}, while the latter has been studied in several nonsupersymmetric backgrounds~\cite{Kruczenski:2003uq,Babington:2003vm} and most successfully realized in~\cite{Sakai:2004cn}. 

In this paper we will lean on the previous results but follow a more phenomenological approach, in the spirit of~\cite{Erlich:2005qh}. There it was shown that a 1-form massless field on the gravity side stands as the natural candidate to study processes involving the QCD conserved current $J_{\mu}={\bar{q}}\gamma_{\mu}q$. Likewise, we argue that a 2-form field, due to its antisymmetric nature, is to be naturally associated with the tensor current $J_{\mu\nu}={\bar{q}}\sigma_{\mu\nu}q$. In the following we will show that this assignment leads to consistent results. However, the first thing to realize is that, as can be inferred from Eq.~(\ref{ads}), we will be dealing with massive 2-form fields. This is consistent with the fact that the tensor current is not conserved in QCD but implies that, in contrast to the vector case, all degrees of freedom of the 2-form will be physical. As we will discuss later on in detail, this will have far-reaching phenomenological consequences. 

Throughout the paper we will be working with a minimal action for 1- and 2-forms, namely that containing only the five-dimensional kinetic terms. Therefore, there will be no coupling between vector and antisymmetric tensors in the gravity side, and it is not clear {\emph{a priori}} how the mixed correlator $\Pi_{VT}$ can be generated from the holographic point of view. We will see that holography provides a very natural scenario for this interplay between QCD currents even starting with uncoupled $p$-forms in five dimensions. At the same time, mixed correlators such as $\Pi_{VT}$ turn out to be particularly sensitive to chiral symmetry breaking effects. However, such effects are more subtle than in correlators like $\Pi_{AA}$ or $\Pi_{AP}$, which contain the pion pole. None of the correlators considered in this work contains the pion pole, and yet they signal the breaking of chiral symmetry. Our aim in this paper is to stick to the simplest five-dimensional Lagrangian and explore where and when the correlators require chiral symmetry breaking effects to appear. As is well-known, conformal symmetry needs to be broken to account for chiral symmetry breaking, typically through nontrivial boundary conditions in the infrared. We show that such infrared boundary conditions are strongly constrained, and therefore correlators with tensor currents turn out to be an ideal theoretical laboratory to understand how to implement chiral symmetry breaking in AdS/CFT. 

However, within the set of correlators considered, not all the requested chiral symmetry breaking effects can be introduced through infrared boundary conditions: changes in the bulk are also needed. In this paper we show that the impact of these changes can be effectively described by slightly modifying the conventional AdS/CFT prescription for correlators. 

This paper is organized as follows: in Sec.~\ref{mathe} we will apply the AdS/CFT prescription to 2-form fields in AdS space with a compact fifth dimension. In Sec.~\ref{compar} we compare the results obtained with the ones from 1-forms, paying special attention to the role played by gauge invariance. Section~\ref{pheno} is devoted to the phenomenological analysis: correlators are defined and the issue of boundary conditions is discussed. In Sec.~\ref{num} we discuss the matching to QCD and the low energy limit of the correlators, providing predictions for low energy parameters and assessing issues like lowest meson dominance. Conclusions are given in Sec.~\ref{conclu}. An interesting result of our analysis is that, quite generally, consistency with generic properties of QCD precludes the presence of dilaton fields in the action. This implies that in order to achieve Regge behavior in holographic QCD, the introduction of a dilaton field does not provide a consistent picture. This is discussed in the Appendix.


\section{2-forms in AdS/QCD}\label{mathe}
The objects we are interested in are generic $p$-form fields in the supergravity side, which will be described by the following action:  
\begin{equation}\label{nform}
S=\kappa\int_{AdS_5}{\mathrm{Tr}}\left[dH\wedge\,^*dH+m^2 H\wedge\,^*H\right]~,
\end{equation}
which satisfies $d ^*H=0$. The integral above is over $(4+1)$ dimensions, and we will adopt the setting where the fifth dimension is finite and delimited by two boundary branes at $y=\epsilon\to 0$ and $y=y_m$, $y$ being the fifth dimension coordinate. Our conventions for the AdS metric are the usual ones:
\begin{equation}
ds^2=g_{MN}dx^Mdx^N=\frac{1}{y^2}(-dy^2+\eta_{\mu\nu}dx^{\mu}dx^{\nu})~,
\end{equation}
with $\eta_{\mu\nu}$ mostly negative. We also include an overall factor $\kappa$ in the action, which in principle could be determined from a more fundamental string theory. 

In this paper we will be mainly interested in 2-forms, {\it{i.e.}}, $H=H_{MN}dx^{M}dx^{N}$, $H_{MN}=H_{MN}^{(a)}\frac{\lambda^a}{2}$. According to the AdS/CFT correspondence, the bulk to boundary propagator of every field in the gravity side corresponds to a current in the gauge theory side. As discussed in the Introduction, a natural candidate for a rank two antisymmetric current in QCD is $J_{\mu\nu}={\bar{q}}\sigma_{\mu\nu}q$, and therefore we will identify $H_{\mu\nu}(\epsilon)$ as the source of $J_{\mu\nu}$. It is well-known that in QCD $J_{\mu\nu}$ can generate $1^{+-}$ mesons through
\begin{equation}
\langle 0|\,J_{\mu\nu}|\,b_n(p,\lambda)\rangle=i f_{Bn}\varepsilon_{\mu\nu\eta\rho}\epsilon^{\eta}_{(\lambda)}\,p^{\rho}~,
\end{equation}
a representative candidate being the $b_1(1235)$ state. However, $J_{\mu\nu}$ can also generate $1^{--}$ mesons through a nonzero $f_{Vn}^{\perp}$, defined as
\begin{equation}
\langle 0|\,J_{\mu\nu}|\,\rho_n(p,\lambda)\rangle=i f_{Vn}^{\perp}(\epsilon_{\mu}^{(\lambda)}\,p_{\nu}-\epsilon_{\nu}^{(\lambda)}\,p_{\mu})~.
\end{equation}
One can understand this duplicity of states in the following way: $\sigma_{\mu\nu}$ has 6 degrees of freedom, which can describe two massive spin-1 states. Since $\sigma_{\mu\nu}\gamma_5=\frac{i}{2}\epsilon_{\mu\nu\lambda\rho}\sigma^{\lambda\rho}$, the states should have opposite parity. Note the analogy with the electric and magnetic fields inside the Faraday tensor.

Since $1^{--}$ states can also be generated by $J_{\mu}={\bar{q}}\gamma_{\mu}q$, in order to be consistent we should also include 1-forms in our analysis. In the literature it is common to interpret $J_{\mu\nu}$ as the generator of longitudinal $\rho$ mesons, while $J_{\mu}$ creates the conventional (transverse) $\rho$ mesons.

In components the action (\ref{nform}) for 2-form fields reads
\begin{eqnarray}
S_2&=&2\kappa \int d^5x \sqrt{g}\,{\mathrm{Tr}}\left[-\frac{1}{2}\partial_{L}H_{MN}\partial^{M}H^{LN}\right.\nonumber\\
&&\!\!\!\!\!\left.+\frac{1}{4}\partial_{L}H_{MN}\partial^{L}H^{MN}-\frac{m^2}{4}H_{MN}H^{MN}\right]~,
\end{eqnarray}
and the equations of motion become
\begin{equation}\label{genEOM}
\frac{1}{\sqrt{g}}\partial_A\bigg[\sqrt{g}\partial_L H_{MN}\Sigma^{ABCLMN}\bigg]+m^2 H_{MN}g^{BM}g^{CN}=0~,
\end{equation}
where 
\begin{eqnarray}
\Sigma^{ABCLMN}&=&g^{AL}g^{BM}g^{CN}+g^{AM}g^{BN}g^{CL}\nonumber\\
&&-g^{AM}g^{BL}g^{CN}~.
\end{eqnarray}
The consistency condition $d^* H=0$ can be expressed in components as
\begin{equation}
\partial_A\bigg[\sqrt{g}H_{MN}(g^{AM}g^{NS}-g^{AN}g^{MS})\bigg]=0~.
\end{equation}
Since $H_{MN}$ is completely antisymmetric, it has $d(d+1)/2$ degrees of freedom, where $d$ is the space-time dimension. Therefore, through Kaluza-Klein reduction the 10 degrees of freedom of $H_{MN}$ split into a field $H_{\mu\nu}$ with 6 components and another field $H_{5\mu}$ with the remaining 4. Schematically,
\begin{equation}
H_{MN}=\left(\begin{array}{ccc|c}
& & & \\
& H_{\mu\nu} & & H_{5\mu}\\
& & & \\
\hline
& H_{\mu 5} & & 0 \\ 
\end{array}\right)~.
\end{equation}
The equations of motion expressed in terms of the fields $H_{\mu\nu}$ and $H_{\mu 5}$ read
\begin{equation}\label{EOM}
-\partial^{\rho}\partial_{\nu}H_{5\rho}+\left(\Box+\frac{m^2}{y^2}\right) H_{5\nu}=\partial_y\partial^{\rho} H_{\rho\nu}~,
\end{equation}
\begin{eqnarray}\label{EOM2}
&&\!\!\!\!\!\!\!\!\!\!\!\!\partial^{\rho}(\partial_{\nu}H_{\rho \mu}-\partial_{\mu}H_{\rho \nu})-\left(\partial_y^2+\frac{1}{y}\partial_y-\Box-\frac{m^2}{y^2}\right) H_{\mu\nu}\nonumber\\
&=&\frac{1}{y}\partial_y\Big[y(\partial_{\nu} H_{5 \mu}-\partial_{\mu} H_{5 \nu})\Big]~.
\end{eqnarray}  
Similarly, it can be shown that the consistency condition amounts to the following relations
\begin{eqnarray}
\partial^{\mu}H_{\mu5}&=&0~,\nonumber\\
H_{5\nu}-y\partial_yH_{5\nu}+y\partial^{\rho}H_{\rho\nu}&=&0~.
\end{eqnarray}
The previous results can be shown to be in agreement with the ones reported in~\cite{l'Yi:1998eu}.

Let us look first at Eqs.~(\ref{EOM}). Using the consistency relations above, the $H_{\mu\nu}$ field can be removed altogether and one obtains the following differential equation for the $H_{5\mu}$ field:  
\begin{equation}\label{h5}
\left(\partial_y^2-\frac{1}{y}\partial_y-\Box+\frac{1-m^2}{y^2}\right)H_{5\lambda}=0~.
\end{equation}
If the two-form is to be associated with the QCD tensor current $J_{\mu\nu}$, application of Eq.~(\ref{ads}) immediately leads to $m^2=1$. In this case, the previous equation has the solution
\begin{equation}\label{solH}
H_{5\lambda}(q,y)=y\big[bJ_1(qy)+Y_1(qy)\big]H_{5\lambda}^{(0)}~,
\end{equation} 
where $q$ is the conjugate momentum of $x$ and $H_{5\lambda}^{(0)}$ and $b$ are initial conditions to be determined later on. 

We will see in Sec.~\ref{compar} that Eq.~(\ref{h5}) with $m^2=1$ is also the equation of motion for a massless vector field $V_{\mu}$. This coincidence is a direct consequence of dimensional reduction in pure AdS spaces, as we will show in the Appendix, and will be extremely important in the interpretation of $H_{5\mu}$ in terms of QCD currents.

Having solved the equation of motion for $H_{5\mu}$ we can now turn our attention back to Eq.~(\ref{EOM2}) to determine $H_{\mu\nu}$. Using again the consistency conditions, the equation can be cast as 
\begin{equation}\label{solH1}
\left(\partial_y^2+\frac{1}{y}\partial_y-\Box-\frac{m^2}{y^2}\right)H_{\rho\lambda}=\frac{2}{y}(\partial_{\rho} H_{5\lambda}-\partial_{\lambda}H_{5\rho})~.
\end{equation}
Notice that no decoupling has been achieved this time and $H_{\mu\nu}$ depends on $H_{5\mu}$. The general solution can be expressed as
\begin{equation}
H_{\rho\lambda}(q,y)=h_{\rho\lambda}(q,y)+{\bar{h}}_{\rho\lambda}(q,y)~,
\end{equation}
where $h_{\rho\lambda}$ is the homogeneous solution, given by
\begin{equation}
h_{\rho\lambda}(q,y)=\big[b^{\prime}J_1(qy)+Y_1(qy)\big]H_{\rho\lambda}^{(0)}~,
\end{equation}
and the particular solution ${\bar{h}}_{\rho\lambda}$ depends on the field $H_{5\mu}$. Knowing the solution for $H_{5\mu}$ from the previous section, one can plug it in and solve the differential equation. The solution is given by
\begin{eqnarray}
{\bar{h}}_{\rho\lambda}(q,y)&=&H^{(0)}_{5\alpha}\frac{i\pi y^2}{2\epsilon}\frac{q_{\rho}\eta_{\lambda}^{\alpha}-q_{\lambda}\eta_{\rho}^{\alpha}}{bJ_1(q\epsilon)+Y_1(q\epsilon)}\nonumber\\
&&\!\!\!\!\!\!\!\!\!\!\!\!\!\!\!\!\!\!\!\!\!\!\!\!\!\!\!\!\!\!\left[\MeijG[2,2][3,5][\xi][\frac{1}{2}][0,\frac{1}{2},-\frac{1}{2}][0,1,-1,-1,-\frac{1}{2}]\left(\frac{Y_1(\xi)-bJ_1(\xi)}{\sqrt{\pi}}\right)\right.\nonumber\\
&&\!\!\!\!\!\!\!\!\!\!\!\!\!\!\!\!\!\!\!\!\!\!\!\!\!\!\!\!\!\!+b\,Y_1(\xi)\left\{J_1^2(\xi)-J_0(\xi)J_2(\xi)\right\}\nonumber\\
&&\!\!\!\!\!\!\!\!\!\!\!\!\!\!\!\!\!\!\!\!\!\!\!\!\!\!\!\!\!\!-J_1(\xi)\left\{Y_1^2(\xi)-Y_0(\xi)Y_2(\xi)\right\}\Bigg]~,
\end{eqnarray}
where $\xi=qy$ and $G^{2,2}_{3,5}$ is a Meijer G function.

Having solved the equations of motion for both $H_{\mu\nu}$ and $H_{5\mu}$ fields one can now plug the on-shell fields back in the action. Only a boundary term survives, namely
\begin{equation} 
S_2=\frac{\kappa}{4}\int d^4x\Bigg\{\sqrt{g}H_{MN}^{(a)}\partial_PH_{RQ}^{(a)}\Sigma^{RPQM5N}\Bigg\}^{y_{m}}_{\epsilon}~,
\end{equation}
which in terms of $H_{\mu\nu}$ and $H_{5\mu}$ can be expressed as
\begin{equation}\label{corrTT}
S_2=-\frac{\kappa}{4}\int d^4x\big[y H^{\mu\nu(a)}(\partial_yH_{\mu\nu}^{(a)}-\partial_{\mu}H_{5\nu}^{(a)}+\partial_{\nu}H_{5\mu}^{(a)})\big]\Bigg|_{\epsilon}^{y_m}\!\!.
\end{equation}
According to the AdS/CFT dictionary, the previous action is the generating functional for the correlators in the gauge theory side.


\section{Comparison with vector fields}\label{compar}

In order to understand better the significance of the results found in the previous section, it is instructive at this point to work out the action for a vector field $V_M=V_M^{(a)}\frac{\lambda^a}{2}$. Equation~(\ref{nform}) for 1-forms reads, in components,
\begin{eqnarray}\label{1form}
S_1&=&\lambda \int d^5x \sqrt{g}\,\,{\mathrm{Tr}}\left[-\frac{1}{2}F_{MN}F^{MN}+m^2V_NV^N\right]\nonumber\\
&=&-\lambda \int d^5x \sqrt{g}\,\,{\mathrm{Tr}}\Big[\partial_MV_N\partial^MV^N\nonumber\\
&&\,\,\,\,\,\,\,\,\,\,\,\,\,\,\,\,\,\,\,\,\,\,\,\,\,\,\,\,-\partial_MV_N\partial^NV^M+m^2V_NV^N\Big]~.
\end{eqnarray}
If $V_M$ is to be associated with the current $J_{\mu}={\bar{q}}\gamma_{\mu}q$, straightforward application of Eq.~(\ref{ads}) gives that $m^2=0$. This means that $V_M$ is a gauge field in 5-dimensions, and as such a gauge fixing procedure is needed. For instance, one could work in the axial gauge and set $V_5=0$ from the beginning, as is done in \cite{Erlich:2005qh}. We will eventually impose the axial gauge condition, but for the time being it will be useful to keep $m^2\neq 0$ and $V_5\neq 0$ all through the analysis. 

The equations of motion applied to Eq.~(\ref{1form}) read 
\begin{equation}
\frac{1}{\sqrt{g}}\partial_A\Big[\sqrt{g}\partial_M V_{N} (g^{AM}g^{BN}-g^{AN}g^{BM})\Big]=m^2V_Ng^{BN}~,
\end{equation}
while the consistency condition $d ^*V=0$, in components, takes the form
\begin{equation}
\partial_A\Big[\sqrt{g}V_Mg^{MA}\Big]=0~.
\end{equation}
Splitting $V_M$ into $V_{\mu}$ and $V_5$ the equations of motion read
\begin{equation}
\left(\Box-\frac{m^2}{y^2}\right) V_5=\partial_y\partial_{\mu}V^{\mu}~,
\end{equation}
\begin{equation}\label{eqV}
\left(\partial_y^2-\frac{1}{y}\partial_y-\Box+\frac{m^2}{y^2}\right) V_{\mu}=y\partial_y\left(\frac{1}{y}\partial_{\mu}V_5\right)-\partial_{\mu}\partial_{\nu}V^{\nu}~, 
\end{equation}
which can be shown to be in agreement with~\cite{l'Yi:1998eu,Mueck:1998iz}.
 
Similarly to what happened in the 2-form case, the equations of motion for $V_{\mu}$ and $V_5$ are coupled, but using the consistency condition $d^* V=0$, which in components can be expressed as
\begin{equation}\label{consV}
3V_5-y\partial_yV_5+y\partial_{\mu}V^{\mu}=0~,
\end{equation}
they get decoupled for the $V_5$ component. The result is
\begin{equation}\label{sol5}
\left(\partial_y^2-\frac{3}{y}\partial_y-\Box+\frac{3+m^2}{y^2}\right)V_5=0~,
\end{equation}
while Eq.~(\ref{eqV}) takes the form
\begin{equation}\label{solV}
\left(\partial_y^2-\frac{1}{y}\partial_y-\Box+\frac{m^2}{y^2}\right)V_{\mu}=\frac{2}{y}\partial_{\mu}V_5~. 
\end{equation} 
Notice that Eqs.~(\ref{sol5}) and (\ref{solV}) above have the same structure as Eqs.~(\ref{h5}) and (\ref{solH1}). In particular, $V_5$ adds a particular solution to the differential equation for $V_{\mu}$ just like $H_{5\mu}$ does to $H_{\mu\nu}$.

The structure of the resulting action is also very similar for vector and tensor fields. For the former one finds: 
\begin{equation}
S_1=\frac{\lambda}{2}\int d^4x\Bigg\{\sqrt{g}V_{M}^{(a)}\partial_PV_{R}^{(a)}\Big(g^{PM}g^{R5}-g^{P5}g^{RM}\Big)\Bigg\}^{y_{m}}_{\epsilon}~,
\end{equation}
or, in terms of $V_{\mu}$ and $V_5$:
\begin{equation}\label{corrVV}
S_1=\frac{\lambda}{2}\int d^4x\,\frac{1}{y} V^{\mu(a)}(\partial_yV_{\mu}^{(a)}-\partial_{\mu}V_5^{(a)})\Bigg|_{\epsilon}^{y_m}~,
\end{equation}
which indeed has the same structure as Eq.~(\ref{corrTT}).

If we now impose that $m^2=0$, Eq.~(\ref{sol5}) becomes the equation for a massive scalar with $m^2=-3$, {\it{i.e.}}, coupled to a Dirac current $J={\bar{q}}q$~\cite{Gubser:1998bc}. However, the very existence of this field would be in serious conflict with QCD, something that can be most easily seen at the level of the action. Inspection of Eq.~(\ref{corrVV}) reveals the presence of two correlators: a quadratic term in $V_{\mu}$, which generates the vector correlator $\Pi_{VV}$, and a mixed term. The physical interpretation of $V_5$ is {\it{a priori}} unclear, but as a scalar field the mixed coupling between $V_{\mu}$ and $V_5$ should account for the $\Pi_{SV}$ correlator. However, this correlator is experimentally found to be zero from analyses on hadronic tau decays~\cite{Ackerstaff:1998yj}, something that can be understood from QCD in the combined chiral and large-$N_c$ limits. In holography, this cancellation is naturally achieved thanks to gauge invariance: $V_5$ can be cancelled and any contribution to $\Pi^{\mu}_{SV}$ is forbidden, in full compliance with QCD.

Let us investigate the consequences of $V_5=0$ a bit further. The required cancellation of $V_5$ reduces the vector equation to the form
\begin{equation}\label{solVV}
\left(\partial_y^2-\frac{1}{y}\partial_y-\Box\right)V^{\mu}=0~,
\end{equation}
whose solution is 
\begin{equation}
V_{\lambda}(q,y)=y\big[bJ_1(qy)+Y_1(qy)\big]V_{\lambda}^{(0)}~.
\end{equation}
Equation~(\ref{solVV}) is to be compared with Eq.~(\ref{h5}) for the $H_{5\mu}$ field. Notice that indeed when $H_{MN}$ is associated with the QCD tensor current $J_{\mu\nu}$, an immediate consequence is that $H_{5\mu}$ satisfies the equation of a massless vector field. Like $V_5$, $H_{5\mu}$ does not have an obvious interpretation in terms of QCD currents. However, unlike $V_5$, it can not be removed by appealing to gauge symmetry, and therefore it must have a physical effect. 

Direct inspection of Eq.~(\ref{corrTT}) shows that there are two potential correlators. Since $H_{\mu\nu}$ can be associated with the tensor current, the term quadratic in $H_{\mu\nu}$ will naturally generate the tensor correlator $\Pi_{TT}$. The nature of the coupling between $H_{\mu\nu}$ and $H_{5\mu}$ is reminiscent of the one between $V_{\mu}$ and $V_5$ and suggests that one should naturally identify the mixed term with $\Pi_{VT}$. This correlator is zero in a conformal theory of quarks, but sensitive to chiral symmetry breaking effects. Therefore, its nonvanishing signals the breaking of chiral symmetry and is demanded by QCD. As we will see later on, in the large-$N_c$ limit $\Pi_{VT}$ is saturated by the exchange of $\rho$ mesons. Therefore, the fact that $V_{\mu}$ and $H_{5\mu}$ are described by the same equation of motion is crucial to have a viable phenomenological description of $\Pi_{VT}$.

We have thus found a remarkable property of holographic AdS models: $H_{5\mu}$ is the vector generator for $\Pi_{VT}$, just as $V_{\mu}$ was the vector generator for $\Pi_{VV}$. This way holography can account for mixed correlators without actually mixing $p$-forms in the five-dimensional action. For consistency, the only thing one has to make sure is that no double-counting takes place, which is guaranteed by the absence of a quadratic term in $H_{5\mu}$ in the action. Note that this picture crucially depends on the fact that $V_5=0$, which is supported by gauge invariance. Most interestingly, as we show in the Appendix, the picture is consistent only in AdS space without (nontrivial) dilaton fields. Thus, one immediate consequence is that dilaton-based mechanisms to implement linear confinement in vector mesons~\cite{Karch:2006pv} will fail to account for $\Pi_{VT}$.   

Having understood the structure and interplay of vectors and antisymmetric tensors, we can now turn to the study of the phenomenological aspects of its associated correlators.


\section{Phenomenological applications}\label{pheno}
In the following we will concentrate on the charged $SU(2)$ flavor currents $J_{\mu}={\bar{u}}\gamma_{\mu}d$ and $J_{\mu\nu}={\bar{u}}\sigma_{\mu\nu}d$. Alternatively, $J_{\mu}^{a}={\bar{q}}\gamma_{\mu}\lambda^{a}q$ and $J_{\mu\nu}^{a}={\bar{q}}\sigma_{\mu\nu}\lambda^{a}q$ with a flavor spurion $\lambda^{a}=\lambda^1+i\lambda^2$.

We define the tensor correlator as 
\begin{eqnarray}\label{tensor}
\Pi^{TT}_{\mu\nu;\alpha\beta}(q)&=&i\int\mathrm{d}^{4}x\, e^{iq\cdot x}\langle \,0\,|\,T\lbrace\, J_{\mu\nu}(x)\,J_{\alpha\beta}^{\dagger}(0)\,\rbrace |\,0\,\rangle\, \nonumber\\
&=&\Pi_{TT}^-(q^2)F_-^{\mu\nu;\alpha\beta}+\Pi_{TT}^+(q^2)F_+^{\mu\nu;\alpha\beta}~,\nonumber\\ 
\end{eqnarray}
where in the second line we have decomposed it in terms of parity-even and parity-odd scalar invariants. The tensor structures $F_{\mu\nu;\alpha\beta}^{\pm}$ are given by 
\begin{eqnarray}\label{structure}
F_-^{\mu\nu;\alpha\beta}&=& q^{\mu}q^{\beta}\eta^{\nu\alpha}+q^{\nu}q^{\alpha}\eta^{\mu\beta}-q^{\mu}q^{\alpha}\eta^{\nu\beta}-q^{\nu}q^{\beta}\eta^{\mu\alpha}~,\nonumber\\
F_+^{\mu\nu;\alpha\beta}&=& -\,\varepsilon^{\mu\nu\sigma\rho}\,\varepsilon^{\alpha\beta\gamma\tau}\,\eta_{\sigma\gamma}\,q_{\rho}\,q_{\tau}\nonumber\\
&=&F_-^{\mu\nu;\alpha\beta}(q)\,+\,q^{2}\left(\eta^{\mu\alpha}\eta^{\nu\beta}-\eta^{\mu\beta}\eta^{\nu\alpha}\right)~.
\end{eqnarray}
and can be shown to be parity projectors. In the following it will prove convenient to define the correlator $\Pi_{TT}^{\pm}\equiv \Pi_{TT}^+-\Pi_{TT}^-$, which is the simplest order parameter of chiral symmetry breaking in the tensor sector, very similar to its counterpart $\Pi_{LR}\equiv \Pi_{VV}-\Pi_{AA}$ in the vector/axial-vector sector.

We will also consider the vector correlator
\begin{eqnarray}\label{vector}
\Pi^{VV}_{\mu\nu}(q)&=&i\int\mathrm{d}^{4}x\, e^{iq\cdot x}\langle\, 0\,|\,T\lbrace\, J_{\mu}(x)\,J_{\nu}^{\dagger}(0)\,\rbrace |\,0\,\rangle \nonumber\\
&=&(q_{\mu}q_{\nu}-q^2\eta_{\mu\nu})\,\Pi_{VV}(q^2)~,
\end{eqnarray}
and the mixed correlator
\begin{eqnarray}\label{mixed}
\Pi^{VT}_{\mu;\nu\rho}(q)&=&i\int\mathrm{d}^{4}x\, e^{iq\cdot x}\langle\, 0\,|\,T\lbrace\, J_{\mu}(x)\,J_{\nu\rho}^{\dagger}(0)\,\rbrace |\,0\,\rangle \nonumber\\
&=&i\,(q_{\rho}\eta_{\mu\nu}-q_{\nu}\eta_{\mu\rho})\,\Pi_{VT}(q^2)~.
\end{eqnarray}
According to the AdS/CFT correspondence, the currents appearing in the previous set of correlators are in one to one correspondence with on-shell fields evaluated on the UV brane. Therefore, given the action for vector and tensor fields,
\begin{eqnarray}
S&=&\int d^4x \left[\frac{\lambda}{2}\,\left\{\frac{1}{y} V^{\mu(a)}\partial_yV_{\mu}^{(a)}\right\}\right.\nonumber\\
&&\!\!\!\!\!\!\!\!\!-\frac{\kappa}{4}\,\left\{y H^{\mu\nu(a)}(\partial_yH_{\mu\nu}^{(a)}-\partial_{\mu}H_{5\nu}^{(a)}+\partial_{\nu}H_{5\mu}^{(a)})\right\}\Bigg]\Bigg|_{\epsilon}^{y_m}~,\nonumber\\
\end{eqnarray}
one can readily obtain two-point correlators by differentiating with respect to the sources. 

Close to the origin, a generic $p$-form $\phi$ behaves like
\begin{equation}\label{scaling} 
\phi(\epsilon,q)=\epsilon^{d-\Delta-p}\phi^{(0)}(q)+\cdots
\end{equation}
where $\phi^{(0)}$ are identified as gauge theory sources~\cite{Witten:1998qj}.

The previous prescription would however fail to properly account for the known high-energy QCD behavior of both $\Pi_{VT}$ and $\Pi_{TT}^{\pm}$. In particular, according to~(\ref{scaling}), $\Pi_{TT}^{\pm}$ would be predicted to diverge logarithmically, as dictated by conformal invariance. However, QCD predicts $\Pi_{TT}^{\pm}$ to converge instead like $q^{-6}$. This is not so surprising bearing in mind that both $\Pi_{VT}$ and $\Pi_{TT}^{\pm}$ are, after all, pure manifestations of chiral symmetry breaking and hence completely unrelated to conformal invariance. This failure to describe $\Pi_{VT}$ and $\Pi_{TT}^{\pm}$ therefore can be seen as a manifestation that conformal invariance and chiral symmetry breaking are not compatible when it comes to the description of {\emph{direct}} order parameters. The breakdown of the AdS/CFT recipe thus points at the necessity of nonperturbative physics away from the conformal limit.

In what follows we will show that instead the phenomenologically-motivated prescription   
\begin{eqnarray}\label{presc}
V_{\mu}(\epsilon,p)&=&V^{(0)}(p)+\cdots~,\nonumber\\
H_{\mu\nu}(\epsilon,p)&=&H^{(0)}_{\mu\nu}(p)+\cdots~,\nonumber\\
H_{5\mu}(\epsilon,p)&=&H^{(0)}_{5\mu}(p)+\cdots~,
\end{eqnarray}
indeed complies with the fundamental requirements of QCD. Insofar as it is motivated by phenomenology our prescription should be interpreted as an effective {\emph{ad hoc}} way to account for the nonconformal aspects that define $\Pi_{VT}$ and $\Pi_{TT}^{\pm}$, namely chiral symmetry breaking but also more general nonperturbative phenomena. In particular, our prescription should capture the fact that the tensor current $J_{\mu\nu}$, unlike the vector $J_{\mu}$, has a nonperturbative dimension different from the conformal one. This is to be expected given that $J_{\mu\nu}$ is not conserved and therefore $\Delta \neq 3$ at strong coupling.

In the following it will prove convenient to split the fields as $H_{\mu\nu}(p,y)=\bar{H}(p,y)H^{(0)}_{\mu\nu}(p)$, $H_{5\nu}(p,y)=\hat{H}(p,y)H^{(0)}_{5\nu}(p)$ and $V_{\mu}(p,y)=\hat{V}(p,y)V^{(0)}_{\mu}(p)$. Accordingly, the UV boundary conditions in~(\ref{presc}) demand that $\bar{H}(p,\epsilon)=\hat{H}(p,\epsilon)=\hat{V}(p,\epsilon)=1$.

Differentiating twice the action with respect to $V_{\mu}^{(0)}$ one obtains $\Pi_{VV}^{\mu\nu}$. Likewise, $\Pi_{TT}^{\mu\nu;\rho\lambda}$ is obtained by differentiating twice with respect to $H_{\mu\nu}^{(0)}$, while $\Pi_{VT}^{\mu\nu;\lambda}$ requires $H_{\mu\nu}^{(0)}$ and $H_{5\rho}^{(0)}$. The results one finds are  
\begin{eqnarray}
\Pi^{VV}_{\mu\nu}(q)&=&i\frac{\delta^2}{\delta V_{\mu}^{(0)}\delta V_{\nu}^{(0)}}S\nonumber\\
&&\!\!\!\!\!\!\!\!\!\!\!\!\!\!\!\!\!\!\!\!\!\!\!\!\!\!\!=-\frac{2\lambda}{q^2}\Big[\frac{1}{y}{\hat{V}}(q,y)\partial_y{\hat{V}}(q,y)\Big]\Bigg|_{\epsilon}^{y_m}(q_{\mu}q_{\nu}-q^2\eta_{\mu\nu})~,\nonumber\\
\Pi^{VT}_{\mu;\nu\rho}(q)&=&i\frac{\delta^2}{\delta V_{\mu}^{(0)}\delta H_{\nu\rho}^{(0)}}S\nonumber\\
&&\!\!\!\!\!\!\!\!\!\!\!\!\!\!\!\!\!\!\!\!\!\!\!\!\!\!\!=-\frac{\kappa}{2}\Big[y \hat{H}(q,y)\bar{H}(q,y)\Big]\Bigg|_{\epsilon}^{y_m} i\,(q_{\rho}\eta_{\mu\nu}-q_{\nu}\eta_{\mu\rho})~,\nonumber\\
\Pi_{\mu\nu;\lambda\rho}^{TT}(q)&=&i\frac{\delta^2}{\delta H_{\mu\nu}^{(0)}\delta H_{\lambda\rho}^{(0)}}S\nonumber\\
&&\!\!\!\!\!\!\!\!\!\!\!\!\!\!\!\!\!\!\!\!\!\!\!\!\!\!\!=-\frac{\kappa}{2q^2}\Big[y{\bar{H}}(q,y)\partial_y{\bar{H}}(q,y)\Big]\Bigg|_{\epsilon}^{y_m}(\eta_{\mu\lambda}\eta_{\nu\rho}-\eta_{\nu\lambda}\eta_{\mu\rho})~.\nonumber\\
\end{eqnarray}
By comparing with Eqs.~(\ref{tensor}), (\ref{vector}) and (\ref{mixed}) the scalar invariants take the form 
\begin{eqnarray}\label{expr}
\Pi_{VV}(q^2)&=&-\frac{2\lambda}{q^2}\Big[\frac{1}{y}{\hat{V}}(q,y)\partial_y{\hat{V}}(q,y)\Big]\Bigg|_{\epsilon}^{y_m}~,\nonumber\\
\Pi_{VT}(q^2)&=&-\frac{\kappa}{2}\Big[y \hat{H}(q,y)\bar{H}(q,y)\Big]\Bigg|_{\epsilon}^{y_m}~,\nonumber\\
\Pi_{TT}^{\pm}(q^2)&=&-\frac{\kappa}{2q^2}\Big[y{\bar{H}}(q,y)\partial_y{\bar{H}}(q,y)\Big]\Bigg|_{\epsilon}^{y_m}~.\nonumber\\
\end{eqnarray}

Note that because of the structure of the action, only the combination $\Pi_{TT}^{\pm}$, which is proportional to the metric, can be extracted. In principle nothing can be inferred directly from $\Pi_{TT}^+$ and $\Pi_{TT}^-$ alone. This is unlike the vector case, where still only the metric piece is accessible but gauge symmetry fills in the missing longitudinal piece.
 
Let us now discuss the choice of boundary conditions. On the UV brane, the fields have to comply with the prescription of Eq.~(\ref{presc}), {\it{i.e.}}, $\bar{H}(p,\epsilon)=\hat{H}(p,\epsilon)=\hat{V}(p,\epsilon)=1$. On the IR brane there is more freedom, and the usual approach is to choose boundary conditions such that there is no infrared contribution to the correlators~\cite{Erlich:2005qh}. However, this approach is valid unless one is dealing with order parameters of spontaneous chiral symmetry breaking (S$\chi$SB). For such quantities, perturbation theory cancels to all orders and one is left with pure nonperturbative effects, which signal the breaking of chiral symmetry. The physical intuition behind the UV brane is that it encodes conformality, while the IR brane is responsible for mimicking confinement and chiral symmetry breaking. Roughly speaking, the UV brane is perturbative while the IR brane is nonperturbative. The correlators $\Pi_{VT}$ and $\Pi_{TT}^{\pm}$ turn out to be order parameters of chiral symmetry breaking, and therefore one expects its holographic expressions to come mainly from the IR brane. 

Let us check all this by examining the UV and IR brane contributions to the different correlators. Given the solutions for $\hat{H}(q,y)$ and $\bar{H}(q,y)$ found in Sec.~\ref{mathe}, we can work out their behavior close to the UV boundary, with the result 
\begin{widetext}
\begin{eqnarray}\label{large}
\hat{H}(q,y)&=&{\hat{V}}(q,y)=1-\frac{q^2y^2}{4}\left[\pi b+\log\left(\frac{q^2}{\mu^2}\right)\right]+\cdots\nonumber\\
\bar{H}(q,y)&=&\frac{y^2}{\epsilon^2}-\frac{y}{\epsilon}+1+\frac{q^2y^2}{4}\left[\pi b^{\prime}+\log\left(\frac{q^2}{\mu^2}\right)\right]\nonumber\\
&&\!\!\!\!\!\!\!\!\!\!\!\!\!\!\!\!\!\!\!\!\!\!\!\!+i\frac{\pi q}{2}q_{\rho}H^{(0)}_{5\lambda}[H^{(0)}_{\rho\lambda}]^{-1}y^2\left[\log\left(\frac{q^2}{\mu^2}\right)-2\sqrt{\pi}\left\{\MeijG[2,0][1,3][0][\frac{1}{2}][-\frac{1}{2}][0,1,-1]-\MeijG[2,1][2,4][0][\frac{1}{2}][\frac{1}{2},-\frac{1}{2}][0,1,-1,-\frac{1}{2}]  \right\}\right]~,\nonumber\\
\end{eqnarray}
\end{widetext}
where $b$ and $b^{\prime}$ are $q$-dependent functions which depend on the boundary condition on the IR brane ($y=y_m$). Since we want $H_{5\mu}$ and $V_{\mu}$ to describe the same physical states, they should have the same boundary conditions, and therefore the same $b$ parameter.

Now we can plug the previous expressions into Eqs.~(\ref{expr}). For the vector correlator, the result is well-known and reads~\cite{Pomarol:1999ad}
\begin{equation}\label{solcorrV}
\Pi_{VV}(q^2)=-\lambda\left[\log\frac{q^2}{\mu^2}-\pi\frac{Y_0\left(\zeta\right)}{J_0\left(\zeta\right)}\right]~,
\end{equation}
where $\zeta=qy_m$. Since $\Pi_{VV}$ is not an order parameter, the usual strategy is to cancel a potential contribution from the IR brane by choosing suitable boundary conditions. It is easy to see that $\partial_y {\hat{V}}(q,y_m)=0$ precisely satisfies this requirement. This choice fixes $b=-Y_0(\zeta)/J_0(\zeta)$, whose explicit expression has already been used in Eq.~(\ref{solcorrV}). 

Let us now turn our attention to $\Pi_{VT}$ and $\Pi_{TT}^{\pm}$. It is straightforward to realize that for $\Pi_{VT}$ the UV brane contribution identically vanishes, while for $\Pi_{TT}^{\pm}$ one obtains $\kappa/2q^2$, which is certainly satisfactory. If the conventional prescription of Eq.~(\ref{scaling}) were used instead, $\Pi_{VT}$ would go like a constant while $\Pi_{TT}^{\pm}$ would diverge logarithmically, in sheer conflict with QCD. This again expresses the fact that Eq.~(\ref{scaling}) misses the chiral symmetry breaking nature of both correlators. 
 
The previous results have to be in agreement with QCD, in particular for large Euclidean momentum. The expressions from QCD there read
\begin{eqnarray}\label{constraints}
\lim_{q^2\rightarrow \infty}\Pi_{VV}(q^2)&=&-\frac{N_c}{12\pi^2}\log{\left(\frac{-q^2}{\mu^2}\right)}+{\cal{O}}\left(\frac{1}{q^4}\right)~,\nonumber\\
\lim_{q^2\rightarrow \infty}\Pi_{TT}^{+}(q^2)&=&-\frac{N_c}{24\pi^2}\log{\left(\frac{-q^2}{\mu^2}\right)}+{\cal{O}}\left(\frac{1}{q^4}\right)~,\nonumber\\
\lim_{q^2\rightarrow \infty}\Pi_{TT}^{-}(q^2)&=&-\frac{N_c}{24\pi^2}\log{\left(\frac{-q^2}{\mu^2}\right)}+{\cal{O}}\left(\frac{1}{q^4}\right)~,\nonumber\\
\end{eqnarray}
where the logarithmic pieces come from the perturbative quark-gluon loop contribution, while subleading pieces in inverse powers of momentum encode the nonperturbative contributions in the OPE.

On the other hand, $\Pi_{VT}$ is given by 
\begin{equation}\label{shortvt} 
\lim_{q^2\rightarrow \infty}\Pi_{VT}(q^{2})=\frac{2\langle{\bar{\psi}}\psi\rangle}{q^2}+\frac{2g_s}{3}\frac{\langle{\bar{\psi}}\sigma_{\mu\nu}G^{\mu\nu}\psi\rangle}{q^4}+{\cal{O}}\left(\frac{1}{q^6}\right),
\end{equation}
{\it{i.e.}}, it is entirely proportional to nonperturbative chiral symmetry breaking vacuum condensates.

The correlator $\Pi_{TT}^{\pm}$ is also an order parameter, but quite different in nature. It is the difference between two logarithmically divergent correlators and satisfies superconvergent relations similar to the ones discovered by Weinberg for the $V-A$ correlator~\cite{Weinberg:1967kj}. One can show that
\begin{equation}
\lim_{q^2\rightarrow \infty}\Pi_{TT}^{\pm}(q^2) \sim \frac{\langle{\bar{\psi}}\psi\rangle^2}{q^6}~.
\end{equation}

Now we can compare the previous QCD high-energy information to the one coming from holography. In the deep Euclidean, Eq.~(\ref{solcorrV}) becomes a pure logarithm
\begin{equation}
\lim_{q^2\rightarrow -\infty}\Pi_{VV}(q^2)=-\lambda\log\frac{-q^2}{\mu^2}~,
\end{equation}
and upon comparison with the parton model logarithm in QCD, one concludes that
\begin{equation}\label{firstmatching}
\lambda=\frac{N_c}{12\pi^2}~.
\end{equation}
Therefore no IR brane contribution is required. However, for $\Pi_{VT}$ a contribution from the IR brane is needed, otherwise the correlator vanishes identically. Any nontrivial IR brane information will play the role of chiral symmetry breaking effects. Notice that for consistency with the vector case we need to impose that $\partial_y {\hat{H}}(q,y_m)=0$ and therefore only the IR boundary condition for ${\bar{H}}(q,y)$ is left unspecified. There are two natural choices, namely 
\[
\begin{array}{ccc}
(a) & \partial_y{\bar{H}}(q,y_m)=\rho(q) & \quad\mathrm{(Neumann)}~, \\ 
(b) & {\bar{H}}(q,y_m)=\rho(q) & \quad \mathrm{(Dirichlet)}~,
\end{array}
\]
where $\rho(q)$ is an arbitrary analytic function which in general will be momentum-dependent.

The first option leads to 
\begin{equation}
{\bar{H}}(q,y_m)=\frac{2\rho(q)}{q}\frac{J_1(\zeta)}{J_0(\zeta)-J_2(\zeta)}~,
\end{equation}
and can be readily excluded, because $\Pi_{VT}$ would then acquire double poles at $J_0$ and single poles at $J_2$. 

A more viable scenario if offered by a Dirichlet boundary condition. Setting ${\bar{H}}(q,y_m)=\rho(q)$, it is easy to show that
\begin{equation}\label{corrle}
\partial_y{\bar{H}}(q,y_m)=\frac{q}{2}\rho(q)\frac{J_0(\zeta)-J_2(\zeta)}{J_1(\zeta)}~.
\end{equation}
On the other hand, close to the IR brane, ${\hat{H}}$ takes the form
\begin{equation}
{\hat{H}}(q,y_m)=-\frac{1}{2}\pi \zeta\left[Y_1(\zeta)-\frac{J_1(\zeta)Y_0(\zeta)}{J_0(\zeta)}\right]~.
\end{equation}
Using the previous equations it is easy to show that the correlators are expressible as
\begin{eqnarray}\label{VTholo}
\Pi_{VT}(q^2)&=&-\frac{\kappa}{2}y_m{\bar{H}}(q,y_m){\hat{H}}(q,y_m)\nonumber\\
&=&\frac{\kappa\pi}{4}y_m^2\rho(q)\left[Y_1(\zeta)-\frac{J_1(\zeta)Y_0(\zeta)}{J_0(\zeta)}\right]\nonumber\\
\end{eqnarray}
and
\begin{eqnarray}\label{TTholo}
\Pi_{TT}^{\pm}(q^2)&=&\frac{\kappa}{2q^2}\Big[1-y_m{\bar{H}}(q,y_m)\partial_{y}{\bar{H}}(q,y_m)\Big]\nonumber\\
&=&\frac{\kappa}{4q^2}\left[2+\zeta\rho^2(q)\frac{J_2(\zeta)-J_0(\zeta)}{J_1(\zeta)}\right]~.\nonumber\\
\end{eqnarray}
Note that indeed the infrared boundary term $\rho(q)$ triggers chiral symmetry breaking. 
  
One can now compare the previous expressions with the form of the correlators in the large-$N_c$ limit of QCD, which are given by ($\xi_n\equiv f_{Vn}^{\perp}/f_{Vn}$)~\cite{Cata:2008zc}
\begin{eqnarray}\label{dispexp}
\Pi_{VV}(q^2) & = & \sum_{n}^{\infty}\frac{f_{Vn}^{2}}{-\,q^{2}+m_{Vn}^{2}}\,,\nonumber \\
\Pi_{TT}^{+}(q^{2}) & = & \sum_{n}^{\infty}\frac{f_{Bn}^{2}}{-\,q^{2}+m_{Bn}^{2}}\,+\,\frac{\Lambda_3}{q^2}\,,\nonumber\\
\Pi_{TT}^{-}(q^{2}) & = & \sum_{n}^{\infty}\xi_{n}^{2}\frac{f_{Vn}^{2}}{-\,q^{2}+m_{Vn}^{2}}\,-\,\frac{\Lambda_3}{q^2}\,,\nonumber \\
\Pi_{VT}(q^2) & = & \sum_{n}^{\infty}\xi_{n}\frac{f_{Vn}^{2}\,m_{Vn}}{-\,q^{2}+m_{Vn}^{2}}\, ,
\end{eqnarray}
with decay constants defined as 
\begin{eqnarray}
\langle 0 |\,J_{\mu}|\,\rho_n(p,\lambda)\rangle&=&f_{Vn}m_{Vn}\epsilon_{\mu}^{(\lambda)}~,\nonumber\\
\langle 0|\,J_{\mu\nu}|\,\rho_n(p,\lambda)\rangle&=&i f_{Vn}^{\perp}(\epsilon_{\mu}^{(\lambda)}\,p_{\nu}-\epsilon_{\nu}^{(\lambda)}\,p_{\mu})~,\nonumber\\
\langle 0|\,J_{\mu\nu}|\,b_n(p,\lambda)\rangle&=&i f_{Bn}\varepsilon_{\mu\nu\eta\rho}\epsilon^{\eta}_{(\lambda)}\,p^{\rho}~.
\end{eqnarray}
Notice that the infrared Dirichlet boundary condition for ${\bar{H}}$ guarantees that $\Pi_{VT}$ has simple poles located at the zeros of $J_0(\zeta)$, exactly like $\Pi_{VV}$ does. Therefore we see that the interpretation of such poles as $\rho$ meson masses is consistent.

Note also that the presence of a nontrivial boundary condition in the IR brane induces a shift in the poles of $\Pi_{TT}^{\pm}$, which now seat at the zeros of $J_1(\zeta)$. This is especially welcome in order to account for the parity-even $1^{+-}$ mesons, as will be shown in the next section.
  

\section{Numerical Analysis}\label{num}

The strategy we will follow is to match our holographic expressions at high energies to the parton model results from QCD and provide predictions for the particle spectrum, {\it{i.e.}}, masses and decay constants, and also low-energy parameters from Chiral Perturbation Theory. This, in particular, will allow us to examine the issue of lowest meson dominance.
  

\subsection{High-energy matching}

Let us start with the decay constants for the $\rho$ meson. Its expressions are easily obtained from the residues of the correlators as follows:
\begin{eqnarray}\label{decay}
f_{Vn}^2&=&2m_{Vn} {\mathrm{Res}}\Big\{\Pi_{VV}(qy_m)\Big\}\Big|_{q=m_{Vn}}\nonumber\\
&=&\frac{N_c \zeta_{0,n}}{6\pi y_m^2}\frac{Y_0(\zeta_{0,n})}{J_1(\zeta_{0,n})}~,\nonumber\\
f_{Vn}f_{Vn}^{\perp}m_{Vn}&=&2m_{Vn} {\mathrm{Res}}\Big\{\Pi_{VT}(qy_m)\Big\}\Big|_{q=m_{Vn}}\nonumber\\
&=&-\kappa\rho(\zeta_{0,n})\frac{\pi}{2y_m}\zeta_{0,n}^2 Y_0(\zeta_{0,n})~,\nonumber\\
\end{eqnarray}
where $\zeta_{0,n}=m_{Vn}y_m$, {\it{i.e.}}, the $n$-th zero of $J_0$. Note that in the first line we already used Eq.~(\ref{firstmatching}) and therefore matching with QCD has already been ensured for $\Pi_{VV}$. The form of $f_{Vn}^{\perp}$ should then guarantee the right matching for both $\Pi_{TT}^-$ and $\Pi_{VT}$. As was shown in Ref.~\cite{Cata:2008zc}, it is convenient to define the ratio $\xi_n=f_{Vn}^{\perp}/f_{Vn}$. Since only $\rho$ mesons propagate in both $\Pi_{VV}$ and $\Pi_{TT}^-$, matching to the parton model only depends on the ratio between the logarithmic coefficients given in Eq.~(\ref{constraints}). Explicitly, Ref.~\cite{Cata:2008zc} found
\begin{equation}\label{pred}
\lim_{n\to\infty}\xi_n\sim\frac{(-1)^{n-1}}{\sqrt{2}}~,
\end{equation}
where the pattern of sign alternation is a direct consequence of $\Pi_{VT}$ being an order parameter, {\it{i.e.}}, there have to be cancellations to comply with the high-energy falloff of Eq.~(\ref{shortvt}).

It should be emphasized that Eq.~(\ref{pred}) is a generic result and should be fulfilled by any model of large-$N_c$ QCD. Notice however that Eq.~(\ref{pred}) is just an asymptotic statement for large excitation numbers and low and middle energies are therefore beyond its scope. Based on a sum rule analysis, the authors of Ref.~\cite{Cata:2008zc} suggested that the pattern should hold all the way down to low energies. This is also suggested by the lattice results on $\xi_{\rho}$~\cite{Becirevic:2003pn}. The holographic model we are describing here should provide a particular realization of those patterns for all energy scales. 

Combining Eqs.~(\ref{decay}) it is straightforward to obtain   
\begin{equation}\label{asymp}
\xi_n=\frac{f_{Vn}^{\perp}}{f_{Vn}}=-\frac{3}{N_c}\kappa\rho(\zeta_{0,n})\pi^2y_m^2J_1(\zeta_{0,n})~.
\end{equation}
Several comments are in order. First of all, note that since $J_1(\zeta_{0,n})$ is positive for odd $n$ and negative for even $n$, the pattern of sign alternation in Eq.~(\ref{pred}) is automatically fulfilled and is predicted to hold at all energies (we are assuming that $\rho(q)$ is monotonic). Geometrically, $f_{Vn}$ is related with the second derivative of the wavefunction at the origin and the sign pattern can be linked to the number of nodes of the wavefunction. On the contrary, $f_{Vn}^{\perp}$ is linked to the IR brane $\rho(q)$ function and the sign is independent of the number of nodes.

Second, parton model matching imposes stringent constraints on the combination $\kappa\rho(q)$. Notice for instance that $\rho(q)=$ct. is incompatible with QCD: since 
\begin{equation}
\lim_{n\rightarrow \infty} J_1(\zeta_{0,n})=0~,
\end{equation}
this implies that $f_{Vn}^{\perp}$ would eventually go to zero, and no matching to the parton model logarithm could be achieved. Since $J_1(\zeta_{0,n})\sim 1/\sqrt{n}$ and $\zeta_{0,n}\sim n$, one needs the combination $\sqrt{\zeta_{0,n}}J_1(\zeta_{0,n})$ to obtain a finite asymptotic limit. At the same time $\rho(q)$ has to be regular at the origin, and therefore the simplest possibility is to choose
\begin{equation}\label{ansatz}
\rho(q)=\rho_0+\rho_1\sqrt{qy_m}~,
\end{equation}
where $\rho_0$ and $\rho_1$ are dimensionless parameters. With the previous ansatz Eq.~(\ref{asymp}) becomes
\begin{equation}
\lim_{n\rightarrow\infty}|\xi_n|=\gamma_{\infty}\kappa\rho_1\frac{3\pi^2}{N_c}y_m^{2}=\frac{1}{\sqrt{2}}~,
\end{equation}
where $\gamma_{\infty}$ is defined as 
\begin{equation}
\lim_{n\rightarrow \infty}\sqrt{\zeta_{0,n}}J_1(\zeta_{0,n})\equiv \lim_{n\rightarrow \infty}\gamma_n=0.797885\equiv \gamma_{\infty}~.
\end{equation} 
Therefore, $\kappa$ (actually the combination $\kappa\rho_1$) is predicted to be
\begin{equation}\label{pred1}
|\kappa\rho_1|=\frac{N_c}{3\sqrt{2}\pi^2\gamma_{\infty} y_m^{2}}~.
\end{equation}   

We now turn our attention to the decay constants for the $1^{+-}$ states. Recall that the holographic prescription only gives information on the combination $\Pi_{TT}^{\pm}$. Nevertheless, since $\Pi_{TT}^+=\Pi_{TT}^{\pm}+\Pi_{TT}^-$, it is straightforward to write down the expression for the decay constants. Note however that the previous equation implies the existence of $1^{+-}$ states as poles of {\emph{both}} $J_0$ and $J_1$. In the following we will denote them by $m_{Bn}^{(0)}$ and $m_{Bn}^{(1)}$ respectively. The expressions for the associated decay constants $f_{Bn}^{(0)}$ and $f_{Bn}^{(1)}$ are
\begin{eqnarray}
\left[f_{Bn}^{(1)}\right]^2&=&2m_{Bn}^{(1)}{\mathrm{Res}}\Big\{\Pi_{TT}^{\pm}(q^2)\Big\}\Big|_{q=m_{Bn}^{(1)}}~,\nonumber\\
\left[f_{Bn}^{(0)}\right]^2&=&2m_{Vn}{\mathrm{Res}}\Big\{\frac{(f_{Vn}^{\perp})^2}{q^2-m_{Vn}^2}\Big\}\Big|_{q=m_{Vn}}~,
\end{eqnarray}
where everything is known on the right-hand side. By inspection, the residues for $J_0$ are exactly $f_{Bn}^{(0)}=f_{Vn}^{\perp}$, and therefore $m_{Bn}^{(0)}=m_{Vn}$. In other words, there is a degeneracy of states with opposite parity. These states are therefore responsible for the parton model logarithms found in both $\Pi_{TT}^+$ and $\Pi_{TT}^-$. Therefore, the remaining states $m_{Bn}^{(1)}$ should have residues in such a way that $\Pi_{TT}^{\pm}$ does not acquire logarithms.

The spectrum of $b_n$ particles is therefore predicted to be doubly dense with respect to the $\rho_n$ meson one. Half the spectrum is completely degenerate with vector mesons and the rest has no impact at the partonic level (at most, it can contribute to higher dimension OPE condensates). Explicitly, one finds
\begin{equation}
\left[f_{Bn}^{(1)}\right]^2=\kappa[\rho(\zeta_{1,n})]^2~,
\end{equation} 
where $\zeta_{1,n}$ are the zeros of $J_1$.

This pattern for the spectrum of $\Pi_{TT}$ is a genuine prediction of holography. This is the only consistent way to take into account: (a) the logarithmic divergences of both $\Pi_{TT}^+$ and $\Pi_{TT}^-$ while (b) their difference $\Pi_{TT}^{\pm}$ being an order parameter of chiral symmetry breaking.

The spectrum therefore looks like
\begin{eqnarray}
m_{Vn}=m_{Bn}^{(0)}&=&\frac{\zeta_{0,n}}{y_m}~,\nonumber\\
m_{Bn}^{(1)}&=&\frac{\zeta_{1,n}}{y_m}~.
\end{eqnarray}
Setting $m_{\rho}=775$ MeV, one finds that $y_m=3.103$ GeV$^{-1}$. The first states for both families are then predicted to be
\begin{equation}
\begin{array}{cc}
m_{\rho}= 775~{\mathrm{MeV}}~; & m_{b1}= 1235~{\mathrm{MeV}}~,\\
m_{\rho^{\prime}}= 1779~{\mathrm{MeV}}~; & m_{b1^{\prime}}=2261~{\mathrm{MeV}}~,\\
m_{\rho^{\prime\prime}}=2789~{\mathrm{MeV}}~;&m_{b1^{\prime\prime}}=3279~{\mathrm{MeV}}~.
\end{array}
\end{equation}
Notice that the $\rho_n-B_n$ splitting is entirely determined by the distance between the zeros of $J_0$ and $J_1$, and that $b_1$ is in excellent agreement with the experimental value, $m_{b1}=1229\pm3$ MeV~\cite{Amsler:2008}. Interestingly, the same splitting is found between $\rho_n$ and axial $1^{++}$ states when chiral symmetry is introduced through boundary conditions~\cite{Hirn:2005nr}. Therefore, it seems that holography predicts $1^{++}$ and $1^{+-}$ states to be degenerate. This is not unnatural taking into account that both states are parity partners of the $\rho$ meson in $\Pi_{LR}$ and $\Pi_{TT}^{\pm}$. Unfortunately, this prediction is hard to test: the PDG reports a promising $m_{b1}=1229\pm3$ MeV and $m_{a1}=1230\pm40$ MeV, but while the $b_1$ can be considered a narrow state ($\Gamma_{b1}=150$ MeV $\sim \Gamma_{\rho}$), the $a_1$ is much broader ($\Gamma_{a1}\sim 250-600$ MeV). This is obviously due to the fact that their main decay channels are $b_1\to\omega\pi$ and $a_1\to 3\pi$, and therefore $b_1$ has a strong phase space suppression.    
  

\subsection{Low-energy predictions}

In the previous section we have fixed $\kappa\rho_1$ from requirements at high energies. At this point the only parameter left is $\rho_0$, which is intrinsically a low-energy quantity. In order to estimate its magnitude we can use the value for $\xi_{\rho}$, which has been determined in lattice simulations to be $\xi_{\rho}\sim 0.72$~\cite{Becirevic:2003pn}. Using Eq.~(\ref{pred1}), $\xi_{\rho}$ can be expressed as
\begin{equation}
\xi_{\rho}=\frac{1}{\sqrt{2}}\frac{\gamma_1}{\gamma_{\infty}}-\frac{3}{N_c}\kappa\rho_0\pi^2y_m^2J_1(\zeta_{0,1})~.
\end{equation}
The first term above already gives $\xi_{\rho}\sim 0.713$ and therefore $\rho_0\sim 0$ to a very good approximation (actually, if it not zero altogether, it has to be unnaturally fine-tuned: $\rho_0/\rho_1\sim 0.04$). In the following we will assume that $\rho_0=0$, which leads to simplifications in the expressions we have found so far. In particular, 
\begin{equation}
\xi_n=\frac{1}{\sqrt{2}}\frac{\gamma_n}{\gamma_{\infty}}~,
\end{equation} 
\begin{center}
\begin{figure}
\includegraphics[width=3.0in]{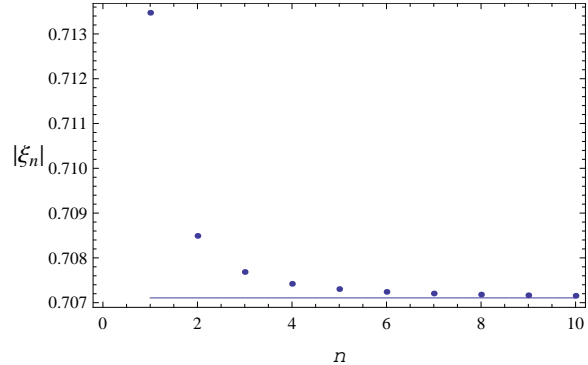}
\caption{$|\xi_n|$ for the first 10 resonance states (blue dots) compared to its asymptotic value (solid line).}\label{xin}
\end{figure}
\end{center}
whose values are plotted in Fig.~\ref{xin}. Notice that parton model matching naturally leads to quasiconstant $\xi_n$. 
Furthermore, one can write
\begin{equation}
\left[f_{Bn}^{(1)}\right]^2=\frac{\rho_1\zeta_{1,n}}{3\sqrt{2}\pi^2\gamma_{\infty}y_m^2}N_c~,
\end{equation}
where $\rho_1$ is expected to be a constant of ${\cal{O}}(1)$.

Let us now turn our attention to the low-energy parameters appearing in the correlators. At low energies, $\Pi_{TT}^{\pm}$ and $\Pi_{VT}$ are parametrized by 
\begin{eqnarray}
\lim_{q^2\rightarrow 0}\Pi_{VT}(q^2)&=&-\chi_0\langle{\bar{\psi}}\psi\rangle+\cdots~,\nonumber\\
\lim_{q^2\rightarrow 0}\Pi_{TT}^{\pm}(q^2)&=&-\frac{\chi_T\langle{\bar{\psi}}\psi\rangle}{q^2}+\cdots~,
\end{eqnarray}
where $\chi_0$ and $\chi_T$ are interpreted as magnetic susceptibilities. Such susceptibilities are low-energy parameters akin to the Gasser-Leutwyler $L_i$ parameters. Its realization as low-energy parameters in a chiral Lagrangian can be found in \cite{Cata:2007ns}.

In order to obtain a prediction for $\chi_0$ in our holographic model, one needs to take the low-energy limit of Eq.~(\ref{VTholo}). Using that 
\begin{eqnarray}
\lim_{q^2\rightarrow 0}{\hat{H}}(q,y_m)&=&1~,\nonumber\\ 
\lim_{q^2\rightarrow 0}{\bar{H}}(q,y_m)&=&\rho_0~,\nonumber\\
\lim_{q^2\rightarrow 0}\partial_y{\bar{H}}(q,y_m)&=&\frac{\rho_0}{y_m}~,  
\end{eqnarray}
one finds the simple expression:
\begin{equation}\label{constLE}
\frac{\kappa}{2}y_m\rho_0=\chi_0\langle{\bar{\psi}}\psi\rangle~.
\end{equation}
In order to determine $\chi_T$ one needs to extract the residue at the origin for $\Pi_{TT}^{\pm}$. Notice that indeed besides the poles at $J_1$, $\Pi_{TT}^{\pm}$ in Eq.~(\ref{TTholo}) has an extra pole at the origin, which matches the $1/q^2$ term present in the large-$N_c$ representation of correlators given in Eqs.~(\ref{dispexp}). Extraction of the residue leads to
\begin{equation}\label{constLE2}
-\frac{\kappa}{3}=\chi_T\langle{\bar{\psi}}\psi\rangle~,
\end{equation}
so that knowledge of $\kappa$ could lead to a prediction for $\chi_T$. For instance, one could estimate it using the sum rule value for the $b_1(1235)$ meson, $f_{b1}\simeq 180$ MeV~\cite{Ball:2006eu}. Then
\begin{equation}
\kappa\simeq(101\,{\mathrm{MeV}})^2;\qquad \rho_1\simeq0.91~,
\end{equation}
and
\begin{equation}
\chi_T\simeq 0.22\,{\mathrm{GeV}}^{-1}~.
\end{equation}

An interesting thing to notice is that, since there is no such $1/q^2$ term for $\Pi_{TT}^{\pm}$ at high energies, there has to be a cancellation between $\chi_T$ and the resonance contribution. The situation is qualitatively similar to what happens with the pion pole in $\Pi_{LR}$, where in order to avoid unwanted $1/q^2$ terms in the OPE one has to impose that
\begin{equation}
\int_0^{\infty}\frac{dt}{t}\frac{1}{\pi}{\mathrm{Im}}\Pi_{LR}(t)=0~,
\end{equation}
which is the celebrated first Weinberg sum rule~\cite{Weinberg:1967kj}. Quite commonly, especially in sum rule applications, the previous expression is cast as
\begin{equation}
\sum_{n}^{N_V}f_{Vn}^2-\sum_{n}^{N_A}f_{An}^2-f_{\pi}^2=0~.
\end{equation}
Note however that the previous expression is valid only if the sum over resonances commutes with the large-$q^2$ expansion. In general this is not a valid manipulation, and certainly not in the limit of large number of colors~\cite{Golterman:2002mi}. 

A similar superconvergence relation holds for the tensor case, and can be written as
\begin{equation}
\int_0^{\infty}\frac{dt}{t}\frac{1}{\pi}{\mathrm{Im}}\Pi_{TT}^{\pm}(t)=0~,
\end{equation}
which, if the spectrum were finite, could be expressed as 
\begin{equation}
\sum_{n}^{N_B}f_{Bn}^2-\sum_{n}^{N_V}\xi_n^2f_{Vn}^2-2\Lambda_3=0~,
\end{equation}
with $2\Lambda_3=-\chi_T\langle {\bar{\psi}}\psi\rangle$.

   
\subsection{A note on lowest meson dominance (LMD)}

Note that $\rho_0=0$, inferred in the previous section from lattice results on $\xi_{\rho}$, opens an intriguing scenario, namely $\chi_0=0$. Different estimates of $\chi_0$ exist in the literature, from QCD sum rules to analyses based on the triangle anomaly in the VVA correlator~\cite{Rohrwild:2007yt}. Here we will focus on a reassessment of methods based on QCD sum rules, where one works at the level of resonance contributions to $\chi_0$. In the large-$N_c$ limit,
\begin{equation}
\chi_0\langle {\bar{\psi}}\psi\rangle=-\sum_n^{\infty}\frac{f_{Vn}^2}{m_{Vn}}\xi_n~,
\end{equation}
which is well-defined as long as the infinite sum exists. In our case, this is guaranteed by construction.

\begin{center}
\begin{figure}
\includegraphics[width=3.0in]{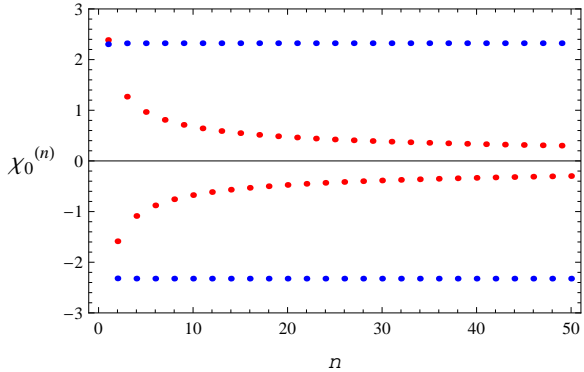}
\caption{Values of $\chi_0^{(n)}$ for the two different scenarios discussed in the main text: blue dots with $|\chi_0^{(n)}|\sim 2.3$ correspond to Eq.~(\ref{first1}) while red ones, decreasing as $1/\sqrt{n}$, correspond to Eq.~(\ref{second2}). LMD is not a good approximation in either case, even though the red dots are a convergent sequence.}\label{xii}
\end{figure}
\end{center}

The contribution of the first few states is (see Fig.~\ref{xii})
\begin{eqnarray}\label{first1}
\chi_0&=&\chi_0^{(1)}+\chi_0^{(2)}+\chi_0^{(3)}+\chi_0^{(4)}+\cdots\nonumber\\
&=&2.300-2.317+2.319-2.320+\cdots~,\nonumber\\
\end{eqnarray} 
and therefore $\rho$-dominance does not hold. Notice that this alternating quasiconstant pattern was to be expected: $\xi_n$ is quasiconstant while $f_{Vn}^2\sim m_{Vn}\sim n$ and therefore resonance contributions do not get suppressed. The eventual cancellation of $\chi_0$ is expressed as an Abel series. However, if one were to extract $\chi_0$ only knowing the first state, one would conclude that $\chi_0\sim 2.3$ GeV$^{-2}$, and the same answer would hold as long as only odd terms were kept in the analysis. $\chi_0$ would then not only appear to be nonzero, but LMD would seem to be satisfied to a high degree.

Failure of LMD in $\Pi_{VT}$ therefore seems to be related to the logarithmic ultraviolet behavior of $\Pi_{TT}^-$. This is at least the conclusion in large-$N_c$ QCD. However, there are reasons to believe that this failure is quite generic. Consider, for the sake of illustration, the case when $\rho(q)=\rho_0$. The interesting thing of this toy scenario is that now $\chi_0\neq 0$, $\xi_n$ vanishes asymptotically and therefore $\Pi_{TT}^-$ is ultraviolet finite. One would expect that this smooth ultraviolet behavior (smoother than QCD) is a much favorable scenario for LMD to hold.

In this new scenario there will still be two sets of $1^{+-}$ particles sitting at the zeros of $J_0$ and $J_1$. The former will still satisfy the degeneracy condition $f_{Bn}^{(0)}=f_{Vn}^{\perp}$, but now
\begin{equation}
\left[f_{Vn}^{\perp}\right]^2=\frac{3\pi^3\zeta_{0,n}}{2N_c}(\kappa\rho_0y_m)^2 J_1(\zeta_{0,n})Y_0(\zeta_{0,n})~,
\end{equation}
whereas the latter are given by
\begin{equation}
\left[f_{Bn}^{(1)}\right]^2=\kappa\rho_0^2~.
\end{equation}
The free parameters of this model are $\kappa$ and $\rho_0$, which can be determined from the lattice value for $\xi_{\rho}$:
\begin{equation}
\xi_{\rho}=-\frac{3}{N_c}\kappa\rho_0\pi^2y_m^2J_1(\zeta_{0,1})\simeq0.72~,
\end{equation}
and the sum rule estimate for the decay constant of the $b_1(1235)$ vector meson. This eventually leads to
\begin{equation}
\kappa\simeq(81\,{\mathrm{MeV}})^2;\qquad \rho_0\simeq-2.22~,
\end{equation}
and using Eq.~(\ref{constLE}) one immediately finds (for typical values of the quark condensate)
\begin{equation}
\chi_0=\frac{\kappa\rho_0y_m}{2\langle{\bar{\psi}}\psi\rangle}\simeq 1.45\, {\mathrm{GeV}}^{-2}~.
\end{equation}
The expression for $\chi_T$ can be found again by extracting the residue of $\Pi_{TT}^{\pm}$ at the origin. The result is
\begin{equation}
\frac{\kappa}{3}(\rho_0^2-1)=\chi_T\langle{\bar{\psi}}\psi\rangle~,
\end{equation}
and therefore
\begin{equation}
\chi_T\simeq -0.55\,{\mathrm{GeV}}^{-1}~.
\end{equation}

Given the value for $\chi_0$, we can again test LMD by computing the contribution of each single state. The result is (see Fig.~\ref{xii})
\begin{eqnarray}\label{second2}
\chi_0&=&\chi_0^{(1)}+\chi_0^{(2)}+\chi_0^{(3)}+\chi_0^{(4)}+\cdots\nonumber\\
&=&2.32-1.54+1.23-1.06+\cdots\nonumber\\
\end{eqnarray} 
Although the first contribution is certainly bigger than the remaining ones, LMD does not apply. The series is actually poorly convergent and for instance one needs 60 iterations to reach a $10\%$ accuracy. Therefore, even in this favorable scenario (certainly more favorable than in the real QCD case) $\rho$-dominance is not a good approximation. 

 
\section{Conclusions}\label{conclu}

We have studied rank two antisymmetric tensors in holographic models of QCD within the AdS/CFT correspondence. Rank two tensors are naturally identified with the tensor QCD current $J_{\mu\nu}={\bar{q}}\sigma_{\mu\nu}q$, which can generate both $1^{--}$ and $1^{+-}$ states. While the latter are genuine states generated by $J_{\mu\nu}$, the former can also be described by the vector current $J_{\mu}={\bar{q}}\gamma_{\mu}q$. Therefore, interference between $J_{\mu\nu}$ and $J_{\mu}$ is to be expected. A convenient way to study this mixing is to examine the set of correlators $\Pi_{VV}$, $\Pi_{VT}$ and $\Pi_{TT}$. A simultaneous analysis of them strongly constraints the spectrum of spin-1 mesons, as already noticed in generic four-dimensional analyses~\cite{Cata:2008zc}. 

One of the advantages of holographic QCD models over four-dimensional ones is that one starts from a Lagrangian formulation. Therefore, one expects the whole set of correlators to show up naturally and provide a self-consistent picture. In this paper we have shown in detail how this interplay is realized in holographic models. 

It is not obvious how a correlator like $\Pi_{VT}$ can be generated starting from the free action for 2-forms in five dimensions. However, we have seen that such mixing arises naturally in the process of compactification. Quite generally, given a $p$-form in five dimensions, a $(p-1)$-form is always dynamically generated by dimensional reduction and realized as the fifth component of the $p$-form. Those dimensionally-reduced fields couple to the original $p$-form and account for mixed correlators. Quite remarkably, they come out with the right mass assignments to be consistent with AdS/CFT: the scalar component of a dimensionally-reduced massless 1-form is a scalar with $m^2=-3$, while the vector component of a dimensionally-reduced massive 2-form is a massless 1-form. Therefore, mixed correlators in holography are natural side-products of four-dimensional compactification. It is worth noting that if the initial $p$-form happens to be massless, then gauge symmetry can be used to remove the mixed correlator. This for instance happens for $\Pi_{SV}$ because $V_5$ can be gauged away. However, $H_{5\mu}$ can not be regarded as a gauge artifact and $\Pi_{VT}$ actually turns out to be nonvanishing, in accord with QCD. This identification, however, crucially depends on the metric being pure AdS without dilaton fields. Therefore, dilaton-based mechanisms in the spirit of~\cite{Karch:2006pv}, designed to display linear confinement in the vector meson spectrum, fail to provide an acceptable phenomenological description of $\Pi_{VT}$.

Another interesting point to notice is that those mixed correlators are order parameters of spontaneous chiral symmetry breaking. As such, their main contribution should naturally come from the infrared brane. Nontrivial boundary conditions on the infrared brane are therefore required, which are somehow related to the existence of a quark condensate. Chiral symmetry breaking is therefore triggered by infrared boundary effects. 

However, we have seen that this is not the full story. The holographic computation of $\Pi_{TT}$ is not free of subtleties. In particular, a direct computation can only be made on the chiral symmetry breaking correlator $\Pi_{TT}^{\pm}$. The AdS/CFT prescription fails to reproduce the right high-energy behavior for this object, implying that conformal invariance needs to be broken beyond infrared boundary terms. Ideally, one should understand how chiral symmetry breaking should be implemented at a Lagrangian level. Instead, in this paper we have made the observation that an {\emph{ad hoc}} modification of the prescription restores agreement with QCD. While this is not fully satisfactory from a formal standpoint, it is certainly enough for phenomenological purposes. 
 
Using this phenomenologically-motivated prescription, we have found that the scenario displayed by holography successfully predicts the presence of $1^{--}$ states in $\Pi_{VV}$ and $\Pi_{VT}$. In particular, the resonance contributions to $\Pi_{VT}$ form an alternate series, in compliance with sum rule analyses and general expectations from QCD~\cite{Cata:2008zc}. $1^{+-}$ states contribute to $\Pi_{TT}$ and are predicted to be doubly dense with respect to $1^{--}$ with masses sitting at the zeros of both $J_0$ and $J_1$. Half of this spectrum is therefore degenerate with $1^{--}$ and the other half with the $1^{++}$ states in the scenario studied in~\cite{Hirn:2005nr}. Additionally, since the decay constants $f_{Vn}^{\perp}$ and $f_{Bn}$ are predicted to be proportional to the infrared boundary value, if chiral symmetry were exact they would vanish altogether. In other words, in the limit of unbroken chiral symmetry, tensor currents do not couple to spin-1 particles.  

Finally, we have assessed lowest meson dominance in the determination of the low-energy parameter $\chi_0$, which is of interest in the hadronic light-by-light scattering contribution to the $(g-2)_{\mu}$. Our results seem to indicate that such assumption is unlikely to work for $\Pi_{VT}$.

\section*{Acknowledgments}
O.~C.~wants to thank the University of Naples for very pleasant stays during the different stages of this work. L.~C.~ and G.~D'A.~ are supported in part by the EU under Contract MTRN-CT-2006-035482 (FLAVIAnet) and by MUIR, Italy, under Project 2005-023102. O.~C.~is supported by the EU under Contract MTRN-CT-2006-035482 (FLAVIAnet) and by MICINN, Spain, under Grants FPA2007-60323
and Consolider-Ingenio 2010 CSD2007-00042 CPAN.\\

\appendix
\section{(Im)possibility of holographic linear confinement with dilaton backgrounds} 
We start from the generic action
\begin{equation}
S=\kappa\int_{AdS_5}e^{-\Phi(y)}{\mathrm{Tr}}\left[dH\wedge\,^*dH+m^2 H\wedge\,^*H\right]~,
\end{equation}
with the general metric 
\begin{equation}
ds^2=g_{MN}dx^Mdx^N=e^{2A(y)}(-dy^2+\eta_{\mu\nu}dx^{\mu}dx^{\nu})~,
\end{equation}  
and allowing for a dilatonic term $\Phi(y)$ with a nontrivial profile in the fifth dimension~\cite{Karch:2006pv}.

The equations of motion for the 2-form field $H_{5\mu}$ can be worked out directly from Eq.~(\ref{genEOM}) with the replacement
\begin{equation}
\sqrt{g}\to \sqrt{\hat{g}}=\sqrt{g}\, e^{-\Phi(y)}~,\qquad \sqrt{g}=e^{5A(y)}~.
\end{equation} 
In this generic case, the consistency condition $d^* H=0$ is expressed as:
\begin{eqnarray}\label{conscond}
\partial^{\mu}H_{\mu5}&=&0~,\nonumber\\
\partial^{\mu}H_{\mu\nu}&=&e^{-A}\partial_y(e^AH_{5\nu})~,
\end{eqnarray}
which can only depend on the metric, as expected.

The first equation of motion, Eq.~(\ref{EOM}), is also dilaton-independent and takes the form
\begin{equation}\label{gene}
(\Box+m^2e^{2A})H_{5\nu}-\partial_{\nu}\partial_{\alpha}H_{5\alpha}-\partial_y\partial^{\alpha}H_{\alpha\nu}=0~,
\end{equation}
from which the equation for $H_{5\mu}$ can be extracted. Since neither Eq.~(\ref{gene}) nor the consistency conditions (\ref{conscond}) depend on the dilaton, it is immediate to conclude that for 2-forms linear confinement cannot be achieved through dilatonic backgrounds. Recall that this is unlike the 1-form component $V_{\mu}$, whose equation of motion does depend on $\Phi(y)$~\cite{Karch:2006pv}. The identification we made of the $H_{5\mu}$ field as a vector field satisfying the same equation as $V_{\mu}$ clearly requires a flat dilaton profile, $\Phi(y)=\phi$.

Let us now look at a different issue, namely the most general subset of metrics consistent with the requirement that $V_{\mu}$ and $H_{5\mu}$ obey the same equation of motion.

Setting $V_5=0$ and using $\partial_{\mu}V^{\mu}=0$, the equation for $V_{\mu}$ is
\begin{equation}
\Box V_{\mu}-e^{-A}\partial_y\Big(e^A\partial_yV_{\mu}\Big)=0~,
\end{equation}
whereas the equation for $H_{5\mu}$, using the consistency conditions, reduces to
\begin{equation}
(\Box+m^2e^{2A}) H_{5\mu}-\partial_y\Big(e^{-A}\partial_y(e^A H_{5\mu})\Big)=0~.
\end{equation}
After some algebra, one can show that the previous equations can be cast as
\begin{eqnarray}
&&\Big(\partial_y^2+e^{-A}\Big(\partial_y e^A\Big)\partial_y-\Box\Big)V_{\mu}=0~,\nonumber\\
&&\Big(\partial_y^2+e^{-A}\Big(\partial_y e^A\Big)\partial_y-\Box\nonumber\\
&&\,\,\,\,\,\,\,\,\,\,\,\,\,\,\,\,-m^2e^{2A}+\partial_y(e^{-A}\partial_ye^A)\Big)H_{5\mu}=0~.\nonumber\\
\end{eqnarray}
Therefore, the wanted metrics satisfy the differential equation
\begin{equation}
\partial_y\Big(e^{-A}\partial_ye^A\Big)=e^{2A}~.
\end{equation}
With the change of variables $\xi=e^{-A}$ one gets
\begin{equation}
(\partial_y\xi)^2-\xi\partial_y^2\xi=1~,
\end{equation}
and the hyperbolic character of the equation is manifest. Its solution is
\begin{equation}
\xi(y)=\frac{\sinh{(ay+b)}}{a}~.
\end{equation}
Notice that pure AdS, {\it{i.e.}}, $e^{2A}=\frac{1}{y^2}$, is recovered when $b=0=a$. For any other choice of parameters, $e^{2A}$ is exponentially damped. 

To summarize, if consistency between $\Pi_{VV}$, $\Pi_{VT}$ and $\Pi_{TT}$ is to be preserved, linear confinement in holographic models of QCD cannot be achieved with dilaton fields.

\end{document}